\documentclass[aps,superscriptaddress,pra,twocolumn,,showpacs]{revtex4-1}
\usepackage{amsfonts, amsmath, amssymb, bm, bbm, graphicx, epsfig, epstopdf}
\usepackage{braket}
\usepackage{dcolumn}
\usepackage{textcomp}
\usepackage{placeins}
\usepackage{hyperref}
\usepackage{color}
\usepackage{soul}
\definecolor{forestgreen}{RGB}{34,139,34}

\usepackage[normalem]{ulem} 
\usepackage{color} 

\begin{document}
	\title{Proposal for transduction between microwave and optical photons using $\mathrm{^{167}Er}$-doped yttrium orthosilicate}
	
\author{Faezeh Kimiaee Asadi} 
\author{Jia-Wei Ji}
\author{Christoph Simon}

\affiliation{Institute for Quantum Science and Technology, and Department of Physics \& Astronomy, University of Calgary, 2500 University Drive NW, Calgary, Alberta T2N 1N4, Canada}

\begin{abstract}
Efficient transduction devices that reversibly convert optical and microwave quantum signals into each other are essential for integrating different technologies. Rare-earth ions in solids, and in particular Erbium ions, with both optical and microwave addressable transitions are promising candidates for designing transducers. We propose a microwave-to-optical quantum transducer scheme based on the dark state protocol in $\mathrm{^{167}Er}$ doped into yttrium orthosilicate (YSO) at zero external magnetic fields. Zero-field operation is beneficial for superconducting resonators
 that can incur extra losses in magnetic fields. By calculating the fidelity and efficiency of the transducer, considering the most important imperfections, we show that an efficient conversion is possible with a high fidelity. We also
investigate the microwave transitions of $\mathrm{^{167}Er}$:YSO that can be used for the transducer protocol. 
  
\end{abstract}

	\maketitle
\section{Introduction}\label{ssec:intro}
Superconducting quantum systems are among the leading candidates for quantum information processing. However, microwave photons, which interact efficiently with superconducting qubits, are not well suited for transmitting quantum information over long distances. This is especially important for designing quantum repeaters that distribute entanglement between remote locations \cite{briegel1998quantum,sangouard2011quantum, kumar2019towards, asadi2018quantum, childress2005fault}. To overcome this problem, the use of microwave-to-optical transducers has been suggested \cite{kurizki2015quantum, lauk2020perspectives}.
There are several mediating systems to host transducers, including atomic ensembles \cite{tu2022high, imamouglu2009cavity, williamson2014magneto, o2014interfacing}, magnons \cite{everts2020ultrastrong, hisatomi2016bidirectional}, electro-optomechanical \cite{hill2012coherent,tian2012adiabatic}, and electro-optical systems \cite{soltani2017efficient}. Solid-state atomic ensembles such as rare-earth (RE) ions \cite{o2014interfacing, williamson2014magneto}, and NV centers \cite{zhao2012scheme,li2017quantum}, in addition to the atomic ensembles in gases \cite{hafezi2012atomic,gard2017microwave,  petrosyan2009reversible}, represent one
of most promising systems for designing transducers as they offer level structures with addressable optical and microwave transitions. On the other hand, compared to atomic gases, solid-state systems are also attractive from the point of view of scalability. Such that, in principle it should be possible to integrate solid-state-based transducers with superconducting
qubits \cite{kumar2019towards}. 

In rare-earth ions doped into a solid, the outer 5s and 5p shells insulate the 4f shell from the crystal environment. As a result, these ions are usually less subject to decoherence at low-temperature. Therefore, rare-earth ion doped crystals are widely used in quantum optics and in particular quantum information storage and signal processing \cite{de2008solid,thiel2011rare, lauritzen2010telecommunication}.

Among rare earth ions with non-zero nuclear spins, Ytterbium ($\mathrm{^{171}Yb}$) with a nuclear spin of $I=1/2$ has the simplest possible hyperfine energy structure which makes the manipulation of spin states quite easy \cite{tiranov2018spectroscopic, kindem2018characterization}. However, it does not have a telecom-wavelength transition. In general, telecom wavelength photons are the best candidates to carry quantum information over long distances due to their minimum absorption in optical fibers. Erbium is a rare-earth ion that offers narrow homogeneous broadening and optical transitions in the telecom window. As a result, several transducer proposals have been developed based on $\mathrm{^{168}Er}$-doped into crystals. In particular, O’Brien et al. \cite{o2014interfacing} proposed the use of a controlled reversible inhomogeneous broadening (CRIB) quantum memory to absorb the incoming pulse in an Er doped Yttrium orthosilicate (YSO) crystal. The absorbed photon will then 
be mapped onto either ground state or optical excitations
depending on the direction of the signal conversion. To improve the efficiency of this protocol, Welinski et al. \cite{welinski2019electron} proposed to use excited state spin levels instead of the ground states as the former are less subject to dephasing mechanisms, and therefore, have a longer coherence time.


On the other hand, there are also some efforts to design transducers based on off-resonant approaches. In this regard, utilizing a Raman-like process, conversion of a microwave signal into the optical field at telecom wavelength has been demostrated in $^{168}$Er:YSO \cite{fernandez2019cavity,fernandez2015coherent}. Most recently, the same group proposed the use of 
an erbium chloride hexahydrate ($^{168}$Er Cl$_3$.6H$_2$O) crystal without disorder 
to design a transducer with enhanced ion densities, but small optical and spin broadening \cite{everts2019microwave}. The use of off-resonant approaches is not limited to rare-earth ions. Using a dark mode of the collective spin excitations, microwave to optical transfer of quantum states has also been discussed for nitrogen-vacancy centers \cite{li2017quantum}.
Most off-resonant schemes are to some extent robust against decoherence mechanisms. 

Erbium has an odd isotope, $\mathrm{^{167}Er}$, with a non-zero nuclear spin of $I=7/2$. A key advantage of using $\mathrm{^{167}Er}$ instead of zero-nuclear spin isotopes of Er is that even at zero magnetic field, $\mathrm{^{167}Er}$:YSO offers around 5 GHz of hyperfine splitting. This is especially important when interacting with superconducting resonators such as superconducting coplanar waveguide cavities that suffer from energy dissipation due to Abrikosov vortex motion in the presence of magnetic fields \cite{song2009microwave}.

Here, utilizing the dark state protocol, we propose the use of $\mathrm{^{167}Er}$ ions doped into YSO for microwave-to-optical transduction in a three-level system at zero external field. YSO is an attractive host crystal because of i) the small nuclear magnetic moments of yttrium ions, and ii) the low isotopic natural abundances of other constituent spins. In addition, Er:YSO has narrow homogeneous and inhomogeneous lines especially at zero external field where the magnetic inequivalence
of atoms vanishes
 \cite{sun2002recent, chen2016coupling}. 
We present a detailed analysis of the dark state transducer protocol and estimate the transfer efficiency and fidelity in Sec. \ref{protocol}.
The implementation of the protocol is discussed in Sec. \ref{impl}.
In this section, using the spin Hamiltonian, we investigate properties of the ground state microwave (MW) transitions of $\mathrm{^{167}Er}$:YSO at zero field, and we list some of the transitions in the GHz regime that can be used for the dark state protocol. Finally, we conclude
and provide an outlook in Sec.\ref{conclusion}

\section{Transduction}\label{protocol}
\subsection{Dark state protocol}
\begin{figure}
\centering
	\includegraphics[width=7cm]{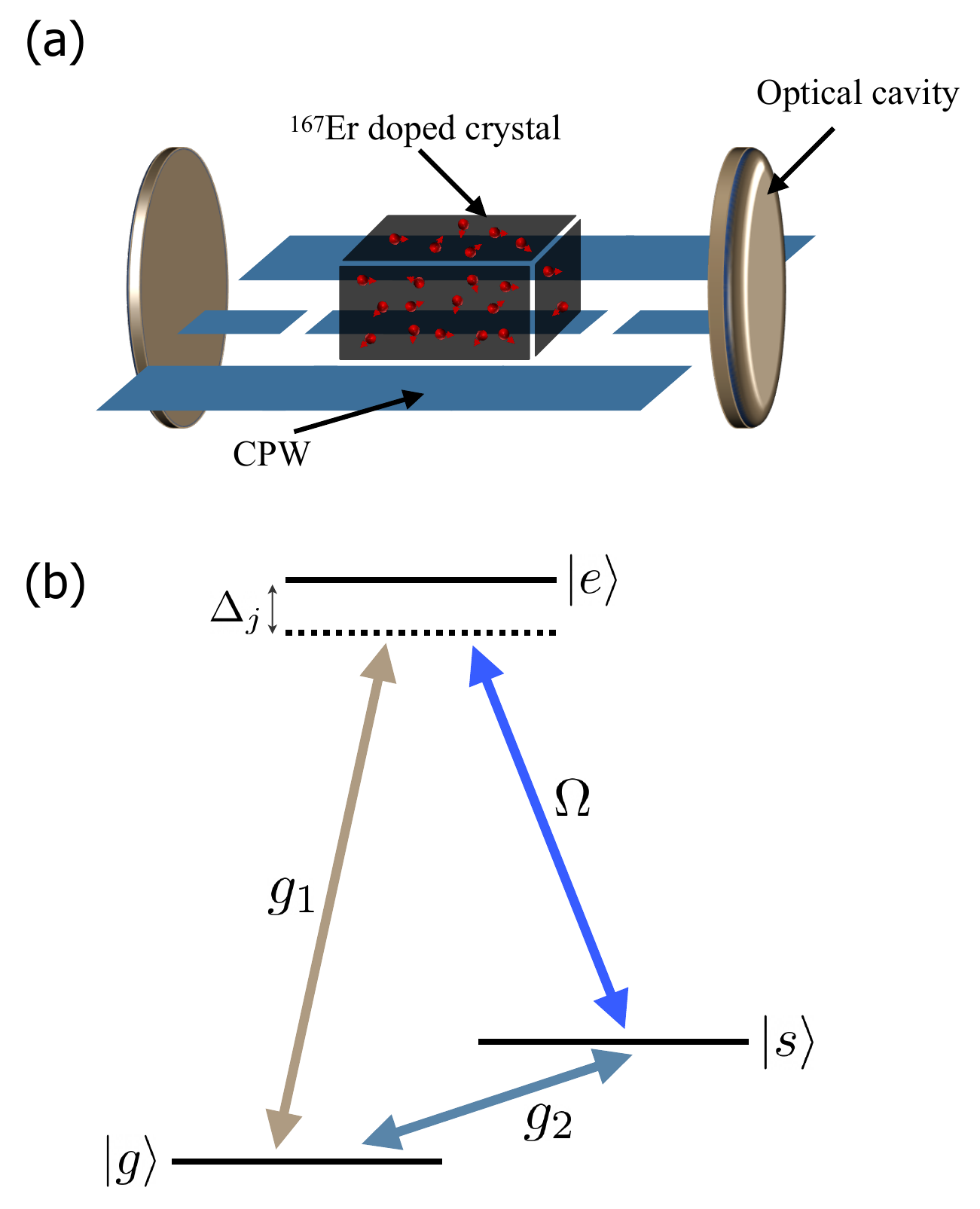}
	\caption{\textbf{(a)} Schematic design of the transducer where the ensemble of $\mathrm{^{167}Er}$ ions doped into YSO is coupled to a microwave superconducting coplanar waveguide and an optical cavity. \textbf{(b)}
	Level diagram for the $j$th ion coupled to an optical cavity and a microwave cavity. This three-level system is driven by a classical field with Rabi frequency $\Omega$, and the transitions $\ket{g}-\ket{e}$ and $\ket{g}-\ket{s}$ are coupled to the optical and microwave photons respectively. The detuning $\Delta_j$ is for the $j$th ion, set to be the same for both transitions.}
	\label{fig: dark}
\end{figure}

Inspired by work on optomechanical systems to transfer quantum states between two different frequencies \cite{wang2012using}, and on four-level nitrogen-vacancy centers in diamond for quantum transduction \cite{li2017quantum}, here we apply the dark state protocol to Er ions with a three-level structure. The main advantage of this protocol is that it is robust against spin decoherence as the collective spin state is only virtually populated during the transfer time. 

Erbium is a Kramers ions, as it has an odd number of 4f electrons, with the ground state
$^4I_{15/2}$ and the lowest excited state of $^4I_{13/2}$. We define the three-level system using the states $\ket{g}$ and $\ket{s}$ from the $^4I_{15/2}$ ground state, and one of the energy levels of the excited state $^4I_{13/2}$ as $\ket{e}$. In Sec.\ref{Microwave-transition}, we provide some examples of energy levels of $\mathrm{^{167}Er}$:YSO that can be used as ground states $\ket{g}$ and $\ket{s}$ at zero external field.

Before we proceed to talk about how this protocol works, we first illustrate the system and the Hamiltonian associated with it. An ensemble of Er ions is placed inside an optical cavity and a microwave superconducting
coplanar waveguide (CPW) cavity. As shown in Fig. \ref{fig: dark}, the optical transition $\ket{g}-\ket{e}$ is coupled to the optical cavity and the transition $\ket{g}-\ket{s}$ is coupled to the microwave cavity, while the transition $\ket{e}-\ket{s}$ is driven by a classical field with Rabi frequency $\Omega$. Here, unlike the scheme of Ref \cite{williamson2014magneto} (where all fields are detuned from the transitions), only the optical fields are detuned. Therefore, as explained in the following paragraphs, we use the dark mode of the collective spin excitations for the conversion and take into account the effect of spin decay and dephasing rates.

For simplicity, we ignore the inhomogeneity in the coupling strength and define two average coupling strengths for ions as $\Tilde{g}_1$ and $\Tilde{g}_2$
\cite{wesenberg2009quantum,amsuss2011cavity,li2017quantum} (for the effect of inhomogeneity in the coupling strength, see Ref\cite{kubo2010strong}). 
The detunings in the optical transition and transition $\ket{e}-\ket{s}$ are set to be the same with $\Delta_j=\omega^j_{eg}-\omega_1=\omega^j_{es}-\omega_\Omega$, where $\omega_1,\omega_\Omega$ are the frequencies for the optical cavity and the classical control field, and the index $j$ indicates the $j$th ion. We introduce the average detuning $\Delta=\Delta_j-\delta_j$ where $\delta_j$ is the inhomogenous broadening for the $j$th spin in the excited state. In the large detuning regime when $|\Delta|\gg |\Omega|, |\delta_j|, |\Tilde{g}_1|, |\Tilde{g}_2|$, the system Hamiltonian can be written as \cite{brion2007adiabatic, james2007effective}:

\begin{equation}
\begin{aligned}
H_{\text{eff}}&=\frac{\Tilde{g}^2_1}{\Delta}\hat{a}^\dagger_1\hat{a}_1\hat{J}_{11}+\frac{\Omega^2}{\Delta}\hat{J}_{22}+(\Tilde{g}_2\hat{a}^\dagger_2+\frac{\Tilde{g}_1\Omega}{\Delta}\hat{a}^\dagger_1)\hat{J}_{12}\\
&+(\Tilde{g}_2\hat{a}_2+\frac{\Tilde{g}_1\Omega}{\Delta}\hat{a}_1)\hat{J}_{21}, 
\end{aligned}
\label{eq:aeff}
\end{equation}
where $\hat{J}_{11}=\sum_{j=1}^{N}{\ket{g}_j\bra{g}}$, $\hat{J}_{22}=\sum_{j=1}^{N}{\ket{s}_j\bra{s}}$, $\hat{J}_{12}=\sum_{j=1}^{N}{\ket{g}_j\bra{s}}$, and $\hat{J}_{21}=\sum_{j=1}^{N}{\ket{s}_j\bra{g}}$ are the collective spin operators. In the low excitation regime, we can apply the Holstein-Primakoff approximation. Then, the above Hamiltonian can be further written as:
\begin{equation}
H_{\text{eff}}=\frac{\Tilde{g}_1\sqrt{N}\Omega}{\Delta}\hat{a}_1\hat{b}^\dagger+\Tilde{g}_2\sqrt{N}\hat{a}_2\hat{b}^\dagger+\text{H.c.},
\label{eq:eff}
\end{equation}
where we ignored the first two terms in Eq. (\ref{eq:aeff}) as they only give us a global energy shift which can be compensated later on. We also used the relations $\hat{J}_{12}\approx \sqrt{N}\hat{b}$ and $\hat{J}_{21}\approx\sqrt{N}\hat{b}^\dagger$ with the operator $\hat{b}$ satisfying the commutation relation $[\hat{b},\hat{b}^\dagger]=1$. Hence, we obtain a Hamiltonian that involves three different bosonic modes. 

Now, let us take several important imperfections into consideration: the optical cavity decay rate $\kappa_1$, the microwave cavity decay rate $\kappa_2$, the collective spin decay rate $\gamma_s$, and the collective spin dephasing rate $\gamma^*_s$. We use the master equation to describe the system dynamics which is given by:
\begin{equation}
\begin{aligned}
\Dot{\hat{\rho}}&=-i[\hat{H}_{\text{eff}},\hat{\rho}]+\kappa_1\mathcal{D}[\hat{a}_1]\hat{\rho}+\kappa_2\mathcal{D}[\hat{a}_2]\hat{\rho}+\gamma_s\mathcal{D}[\hat{b}]\hat{\rho}\\
&+\gamma^*_s\mathcal{D}[\hat{b}^\dagger\hat{b}]\hat{\rho},
\end{aligned}
\label{eq:mas}
\end{equation}
where $\hat{H}_{\text{eff}}$ is given in Eq. (\ref{eq:eff}), and $\mathcal{D}[\hat{A}]\hat{\rho}=\hat{A}\hat{\rho}\hat{A}^\dagger-\hat{A}^\dagger\hat{A}\hat{\rho}/2-\hat{\rho}\hat{A}^\dagger\hat{A}/2$.

Eq. (\ref{eq:eff}) can be fully diagonalized with three distinct eigenmodes, which are $\hat{C}_d=\frac{-G_2\hat{a}_1+G_1\hat{a}_2}{\sqrt{G^2_1+G^2_2}}$, $\hat{C}_{\pm}=1/\sqrt{2}(\frac{G_1\hat{a}_1+G_2\hat{a}_2}{\sqrt{G^2_1+G^2_2}}\pm\hat{b})$ with $G_1=\frac{\Tilde{g}_1\sqrt{N}\Omega}{\Delta}$ and $G_2=\Tilde{g}_2\sqrt{N}$. It is crucial to see that the mode $\hat{C}_d$ is referred to as "dark state" as it decouples from the collective spin mode $\hat{b}$. Here, the basic idea is to modulate parameters $G_1(t)$ and $G_2(t)$ such that for microwave-to-optical transfer at $t=0$, $\hat{C}_d=\hat{a}_2$, and  at $t=t_f$, $\hat{C}_d=-\hat{a}_1$. It has been shown that the optimal modulation can be obtained by setting $G^2_1(t)+G^2_2(t)=G^2$, where $G$ is a constant \cite{vasilev2009optimum,wang2012using}. However, it is quite difficult to modulate the single-photon coupling rate $\Tilde{g}_2$. Thus, here we set $G_1(t)=G \sqrt{1-\text{tanh}(\alpha t)}$ while keeping $G_2(t)=G$ constant with $\alpha$ being the modulation strength parameter. See Appendix. \ref{Optimal-modulation} for more information on the role of $\alpha$.


\subsection{Efficiency and fidelity}\label{Dark-EF}

Let us first define the transduction efficiency and fidelity. Here we focus on a single-photon input. In the scenario where we attempt to convert a single microwave photon to an optical photon, we define the efficiency as
\begin{equation}
\eta=\text{Tr}[\hat{\rho}_f \hat{a}^\dagger_1 \hat{a}_1],
\end{equation}
with $\hat{\rho}_f$ being the final state of the system. 
Here, the final state depends on the protocol time $t_f$. In this work, we choose $t_f=0.5/\alpha$ as a function of $\alpha$. Note that this $t_f$ (which is among many other similar choices) is picked to yield reasonable values of efficiency and can be further optimized. 

\begin{figure}
    \centering
\includegraphics[scale=0.5]{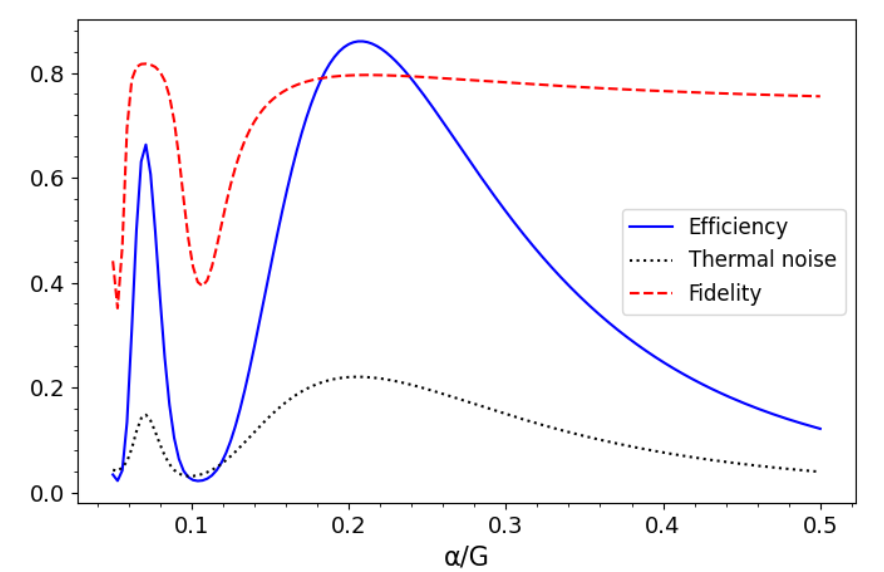}
	\caption{Efficiency, fidelity and thermal noise of the dark-state protocol as a function of $\alpha/G$ for microwave-to-optical transfer.
	Here for the parameters we assume $G/2\pi=10$ MHz, $\kappa_1=0.1 G$, $\kappa_2=0.02 G$, $\gamma_s=0.001 G$, $\omega_{gs}/2\pi=1.33$ GHz, $T=50$ mK, and $\gamma^*_s=0.0358 G$.} 
	\label{fig:F-E}
\end{figure}
Fidelity is often defined as the overlap between the density matrices at the beginning and end of the transfer process. Considering that the differences in mode shape can be corrected by unitary transformations, here we instead focus on the role of noise due to thermal excitations (microwave photons) which is likely to be the most important challenge for quantum network implementations. Therefore, to quantify the fidelity, we use the signal-to-noise ratio
(SNR) and set:
 
\begin{equation}
F_\text{SNR}=\frac{1}{1+\text{SNR}^{-1}}.    
\end{equation}
The SNR takes the form $\text{Tr}[\hat{\rho}_f \hat{a}^\dagger_1 \hat{a}_1]/\text{Tr}[\hat{\rho}'_f \hat{a}^\dagger_1 \hat{a}_1]$ where $\hat{\rho}'_f$ is the final state without any input. Here, the average number of thermal microwave photons is given by  $\bar{n}_{\text{th}}=1/(e^{(\hbar\omega_{gs}/k_B T)}-1)$ where $\omega_{gs}$ is the frequency between two microwave transitions $\ket{g}-\ket{s}$, and $T$ is the system temperature. Note that, in our application the signal is due to collective interference from many atoms, whereas dephasing leads to much weaker (non-collective) background emission. Therefore, we expect the dephasing rate to primarily affect the efficiency but not the fidelity \cite{staudt2007interference}.

In Fig. \ref{fig:F-E}, we set $t_f=0.5/\alpha$ and plot the transduction efficiency, fidelity and noise with respect to different values of $\alpha/G$. If $\alpha$ is too large (close to $G$ or even larger than $G$) the adiabaticity can no longer be maintained as the collective spin mode in the ground state will be occupied. This can degrade the transduction efficiency. On the other hand, if $\alpha$ is too small (close to cavity decay rates), the transduction efficiency will be largely affected by the cavity decay rates. Therefore, we need to optimize this parameter. At around $\alpha/G=0.212$ and $T=50$ mK, the efficiency reaches its maximum value of $0.86$. At this efficiency, the corresponding fidelity is $0.817$. 

One should notice that here the adiabatic transfer is non-ideal as modulation functions $G_1$ and $G_2$ start out equal and end with $G_2$ not much larger than $G_1$. Thus, it is more affected by spin decoherences than the ideal transfer but it still yields reasonably good efficiencies due to the fact that compared to interaction strength $G$, the strengths of spin decoherences are much smaller. On the other hand, one can make the protocol time longer to better keep the transfer close to the ideal case but this will subject the system to more dissipations. Therefore, we can see there is some room left for the full optimization of both $t_f$ and $\alpha$, which could be further explored in the future work. 

Here, we set the dephasing rate as $\gamma_s^\star=2\pi \times 358 $ kHz. It has contribution from both spin inhomogeneous broadening  and decoherence rate. The former is $\gamma_{IB}=2\pi \times 350$ kHz \cite{rakonjac2020long}. For the latter, we used $T_2=12 \mu s$ and estimated the pure dephasing rate ($\gamma^\star$) using the relation $1/T_2=\gamma^\star+\gamma_s/2$ assuming the spin decay rate of $2\pi \times 10$ kHz. We justify the values we used for the transition frequency and coherence time in Sec. \ref{Microwave-transition}.

\begin{figure}
    \centering
     $\hspace{-70mm}\mathbf{(a)}$\\
          \vspace{2mm}
    \includegraphics[scale=0.5]{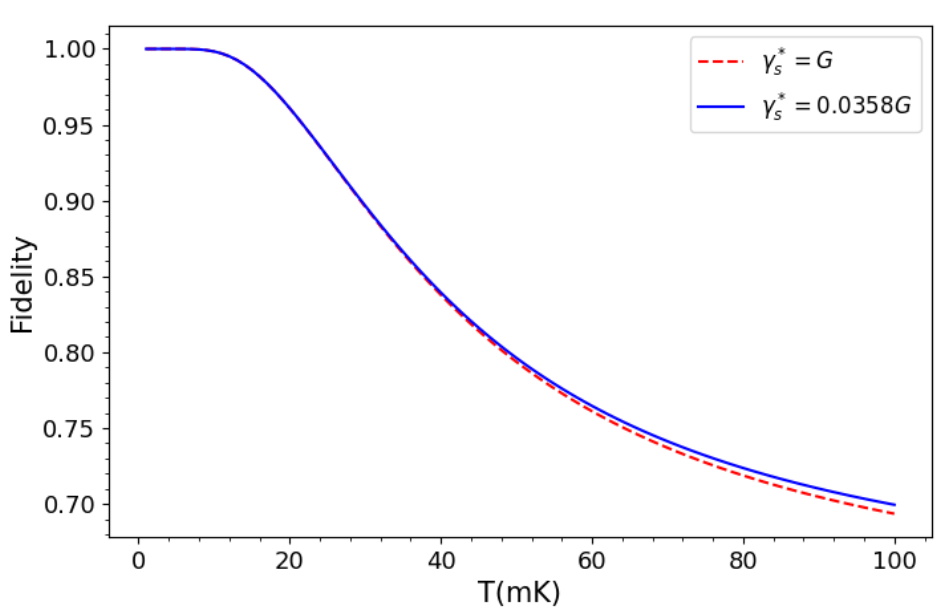}\\
    $\hspace{-70mm}\mathbf{(b)}$\\ \vspace{2mm}
    \includegraphics[scale=0.5]{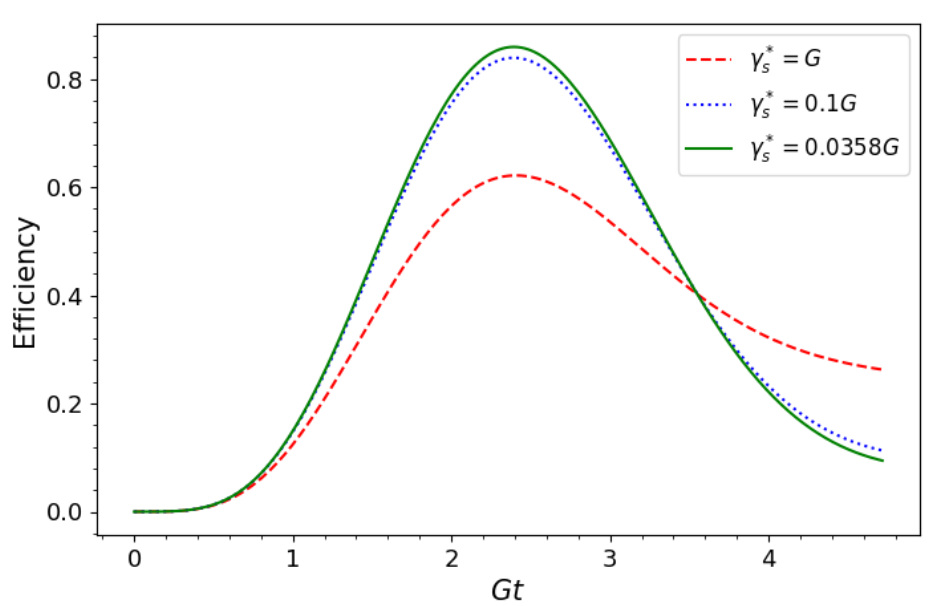}
	\caption{(a). Fidelity as a function of temperature for two different dephasing rates. Here we set $G/2\pi=10$ MHz, $G/\alpha=0.212$, $\kappa_1=0.1 G$, $\kappa_2=0.02 G$, $\gamma_s=0.001 G$, and $\omega_{gs}/2\pi=1.33$ GHz. The whole protocol time is also fixed to be $t_f=0.5/\alpha$ with $\alpha=0.212 G$. 
	(b). Efficiency as a function of time for three different dephasing rates. The other parameters are the same as those used in (a), and the temperature is $T=50$ mK.} \label{fig:Fidelity}
\end{figure}

In Fig. \ref{fig:Fidelity}(a), for a fixed $\alpha/G=0.212$ with the protocol time assumed to be $0.5/\alpha$, we have shown the change in fidelity of the microwave-to-optical transfer with respect to the temperature. By increasing the temperature, the average number of thermal microwave photons increases. As a result, the transfer fidelity decreases. Note that, superconducting qubits require temperatures in the mK range. 
In this figure, we have also shown fidelity changes for two different dephasing rates. 
As expected, the transduction fidelity is robust against the dephasing rates but this figure of merit does not fully capture the robustness of dark-state protocol. Thus, we plot the transduction efficiency with respect to time as shown in \ref{fig:Fidelity}(b) for better illustration. As can be seen, in this protocol the efficiency is also to some extent robust against the dephasing rates. When the dephasing rate $\gamma^*_s=0.1 G$, the efficiency is still above 80$\%$, slightly lower than the one with $\gamma^*_s=0.0358 G$. However, it still has some limits as when the dephasing rate is very large, e.g. $\gamma^*_s=G$, the efficiency is around 60$\%$.


\section{experimental implementation}
\label{impl}
To spin polarize $^{167}\text{Er}$ ions, usually an optical laser sweeps over the range of  $\Delta m_I=1$ (or $\Delta m_I=-1$ depending on the hyperfine level of interest) transitions \cite{stuart2021initialization}. Here $m_I$ indicates the quantum number of the nuclear spin. 
However, in the low-field regime, since the transitions of $\Delta m_I=0$ and $\Delta m_I=\pm1$ are not clearly resolved, $m_I$ is not a good quantum number anymore and only partial spin polarization is possible. It is therefore crucial to operate at ultra-low temperatures. This can help to some extent by freezing out the spin background noise. In addition, for an efficient spin polarization at zero field, both microwave and optical pumping should be employed.

Our system is composed of an ensemble of $\mathrm{^{167}Er}$ ions in an optical and a microwave superconducting
coplanar waveguide (CPW) cavity. We consider a case where the photon to be transferred is initially in the microwave cavity. For the microwave side, the coupling of rare-earth spin ensembles to a microwave cavity has been demonstrated \cite{probst2013anisotropic, tkalvcec2014strong, staudt2012coupling, chen2016coupling}, and the coupling strength of $2\pi \times34$ MHz is reported in Ref. \cite{probst2013anisotropic}. Thus, it is reasonable to assume the collective coupling strength $\Tilde{g}_2\sqrt{N}\sim 2\pi\times 10$ MHz in our scheme. Furthermore, for the CPW cavity, a quality factor $Q\sim 10^6$ is possible to achieve \cite{niemczyk2010circuit, xiang2013hybrid}, thus giving us $\kappa_2\sim 2\pi\times 1.33$ kHz for the $\omega_{gs}/2\pi=1.33$ GHz. In our simulation, for a CPW resonator that is coupled to a crystal at mK temperatures, we assume a higher decay rate of $\kappa_2\sim 2\pi\times 200$ kHz corresponding to the quality factor of $6.6\times 10^3$. 

So far, for Fabry-Pérot cavities, the quality factor of $Q\sim 10^9$ has been realized \cite{xiang2013hybrid,aoki2006observation,goto2010experimental}. Using toroid microcavities, the $Q$ factor of $4\times 10^8$ has also been measured for $1550$ nm wavelength \cite{kippenberg2004demonstration}. In addition, a quality factor exceeding $1.1 \times 10^7$ has been reported in
photonic crystal nanocavities \cite{asano2017photonic}. Here, we consider the decay rate of $\kappa_1=2\pi \times 1$ MHz corresponding to $Q \sim10^8$ for the optical cavity. 
In a single rare-earth ion coupled to a photonic crystal resonator with a small mode volume, the coupling strength of $ 2\pi \times 28.5\pm5.2$ MHz has been reported \cite{zhong2018optically}. Coupling strength can be significantly enhanced in an ensemble of ions. Here we assume $\Tilde{g}_1\sqrt{N}\sim 2\pi\times 500$ MHz, considering that the number of ions can be around $10^3$ in a comparably small volume.


Inhomogeneous broadening in optical transitions is typically $\delta_j\sim 2\pi\times 1$ GHz \cite{thiel2011rare,baldit2005ultraslow}. We take the average optical detuning $\Delta\sim 2\pi\times 10$ GHz, which satisfies that $\Delta\gg\delta_j$. The Rabi frequency is determined by the laser power and beam diameter. In ref \cite{stuart2021initialization}, it has been discussed that by setting the spot size as $10 \mu$m and amplifying the laser using erbium doped fiber amplifier, a Rabi frequency on the order of $2\pi \times 100$ MHz is achievable at few-Kelvin temperatures. Here we set $\Omega\sim 2\pi\times 200$ MHz, which also satisfies $\Omega\ll\Delta$. With all these values, we estimate $G_1=\Tilde{g}_1\sqrt{N}\Omega/\Delta\sim 2\pi\times 10$ MHz. Note that to achieve a Rabi frequency of $O (100)$ MHz at $\sim 100$mK temperatures, we need a smaller spot size (i.e., focused beam) to keep the power low. Otherwise, there would be excess heating in the system. The required spot size is determined by the cryostat's cooling power and the percentage of the absorbed power.

In general, the mode mismatch factor depends on the design of the cavities including mode volumes, and microwave and optical mode functions. For an ensemble of $^{168}\text{Er}$:YSO coupled to a Fabry-Pérot and a loop-gap cavity, the mismatch factor can be as small as $0.0084$ \cite{williamson2014magneto}. Here for simplicity, we assume that the ions are located in the maximum of both cavities. Therefore, we ignore the mode mismatch factor.



\subsection{Microwave transitions of $^{167}\text{Er}$:YSO}

\begin{table*}[t]
	\caption{ Ground state transition frequencies, transition strengths and coherence times  for site 1 of $\mathrm{^{167}Er}$:YSO at zero magnetic field. Energy levels are labeled as 1 - 16 from lowest to highest frequency. Here transition dipole moments are determined  relative to the three orthogonal optical extinction axes defined by $D_1(X)$, $D_2(Y)$, and $b(Z)$.}
	\begin{ruledtabular}
		\begin{tabular}{c  c  c  c  c  c }
			Transition frequency (GHz) &  $d(D_1,D_2,b)$ (GHz/T) & \,\,\,\,\,\, Coherence time ( $\mu$s) \\[0.05cm]
			\hline \\[-0.2cm]
			
			1.33 (7 $\longleftrightarrow$10  )\,\, & (0.48,\,2.05,\,0.59) &12.16\\[0.05cm]

			2.374 (6 $\longleftrightarrow$12  )\,\, & (3.66,\,6.43,\,1.66) &4.56\\[0.05cm]		
			
			2.366 (5 $\longleftrightarrow$11  )\,\, & (2.41,\,9.15,\,0.34) &4.4\\[0.05cm]			
			1.821 (7 $\longleftrightarrow$11  )\,\, & (3.35,\,12.83,\,7.3) &1.61\\[0.05cm]				
			1.304 (8 $\longleftrightarrow$12  )\,\, & (3.52,\,13.01,\,7.51) &1.59\\[0.05cm]			
	
		\end{tabular}
	\end{ruledtabular}
	\label{tab:dipole}
\end{table*}

\label{Microwave-transition}
Erbium  has eight Kramers’ doublets in the ground state
and seven in the excited state. Due to the hyperfine and quadrupole interactions, each doublet is split into the sixteen hyperfine sub-levels. At low temperatures, only the lowest doublet is populated. The effective spin Hamiltonian of the Kramers' ions with non-zero nuclear spin can be written as \cite{abragam2012electron}
\begin{equation}
H_{eff}=\beta_e \bold{B} \cdot \bold{g} \cdot \bold{S} + \bold{I} \cdot \bold{A} \cdot \bold{S}+\bold{I} \cdot \bold{Q} \cdot \bold{I} - \beta_n g_n \bold{B} \cdot \bold{I},\label{H}
\end{equation}
where $\beta_e (\beta_n)$ is the electronic Bohr (nuclear) magneton, $\bold{B}$ is the external magnetic field, $\bold{A}$ is the hyperfine tensor, $\bold{Q}$ is the electric-quadrupole tensor, $\bold{S}$ ($\bold{I}$) is the vector of electronic (nuclear) spin operator, $\bold{g}$ is the $g$ tensor, and $g_n$ is the nuclear g-factor. The first term of the above Hamiltonian describes the electronic Zeeman interaction. The second and third terms describe the hyperfine and electric quadrupole (second-order hyperfine) interactions. Finally, the last term is the nuclear Zeeman interaction.

The spin Hamiltonian parameters have been estimated for $\mathrm{^{167}Er}$:YSO using the electron spin
resonance experiment and crystal field model for the ground \cite{chen2018hyperfine}, and excited \cite{horvath2019extending} states, respectively. Using the spin Hamiltonian formalism, there is a good agreement between the transition frequency estimations and experimental results, i.e., the difference is less than $\sim40$ ($\sim$100) MHz for the ground (excited) states \cite{rakonjac2020long, horvath2019extending} (see Appendix \ref{energy-levels} for the MW transitions in the ground state).  


At zero magnetic
field, the hyperfine
structure of $^{167}\text{Er}$:YSO is split over 5.4 GHz. In Table.\ref{tab:dipole}, we have listed top five transition frequencies in the GHz regime with longest coherence times for zero magnetic field. Here, we calculated the transition dipole moments $d_{D_1,D_2,b}$ between energy levels in the ground hyperfine structure.
To do so, we consider the atom-field interaction Hamiltonian
\begin{equation}
H_{I}=\beta_e \bold{B_{ac}} \cdot \bold{g} \cdot \bold{S}  - \beta_n g_n \bold{B_{ac}} \cdot \bold{I}\label{H},
\end{equation}
where the transition is being driven by an ac magnetic field. Then, to estimate the magnetic dipole
moment of a transition along the direction of the ac field, we use the probability amplitude scheme \cite{scully1999quantum}
\begin{equation}
d_{mn}=\bra{\psi(B)_m}\beta_e  \bold{g} \cdot \bold{S}  - \beta_n g_n \bold{I}\label{H} \ket{\psi(B)_n}.
\end{equation}
In the absence of a magnetic field, electronic and nuclear states are highly mixed. Hence, transition moments are larger than the nuclear magneton.

In rare-earth ions doped crystals, spin flips can occur due to the spin-spin (i.e., spin flip-flop) and spin-lattice relaxations. At low magnetic fields, which is the relevant regime here, spin flip-flops is the governing mechanism. Decreasing the flip-flop rate is possible by reducing the temperature
to polarize the spins. Note that low temperatures is also required for the benefit of the superconducting qubits. Hence, we consider an ensemble of $\mathrm{^{167}Er}$:YSO in sub kelvin temperatures. 
In this case, flipping of nearest neighbour ions is the dominated perturbation mechanism that contributes to the spin decoherence. Considering the distance to the nearest neighbour Erbium and Yttrium ions,  
one can estimate the variance of the magnetic field fluctuation created at the Er site as $\Delta B=26 \mu T$ \cite{thesis}. To estimate the coherence time in Table.\ref{tab:dipole}, we assume the decoherence time occurs on a timescale much longer than the magnetic field fluctuations. In this case, the coherence time is given by \cite{zhong2015optically}
\begin{equation}
\frac{1}{\pi T_2}=S_1\cdot \Delta B+ \Delta B \cdot S_2 \cdot  \Delta B, \label{coherenceT}
\end{equation}
where $S_1$ is the gradient, and  $S_2$ is the curvature of the transition of the interest.
To calculate $S_1$ and $S_2$, we define the Zeeman gradient and curvature Tensor parameters as \cite{mcauslan2012reducing}
\begin{equation}
\begin{aligned}
\nu_i^{mn}(B)&=\frac{\partial\,(\omega_m(B)-\omega_n(B))}{\partial B_i},\\
C_{ij}^{mn}&=\frac{\partial^2(\omega_m(B)-\omega_n(B))}{\partial B_i\partial B_j},
\end{aligned}
\end{equation}
where 
\begin{equation}
\begin{aligned}
&\frac{\partial\,\omega_m(B)}{\partial B_i}  =\bra{\psi(B)_m}\!\zeta_{ij}\!\ket{\psi(B)_m},\\
\frac{\partial^2\omega_m(B)}{\partial B_i\partial B_j} & =\!\!\sum\limits_{m\neq n}\!\!\frac{\bra{\psi(B)_m}\!\zeta_{ik}\!\ket{\psi(B)_n}\!\bra{\psi(B)_n}\!\zeta_{jl}\!\ket{\psi(B)_m}}{\omega_m(B)-\omega_n(B)}.
\end{aligned}
\end{equation}
Here $\ket{\psi(B)}$ is the state of the system, $\omega(B)$ is its corresponding energy, $ \omega_m(B)-\omega_n(B)$ is the frequency difference for the $m \longleftrightarrow n$ transition, and $\zeta_{ij}=\beta_e \bold{g_{ij}}\bold{S_j}-\beta_ng_n \bold{I_i}$ where $\bold{S}$ is the spin operator, and $\bold{I}$ is the nuclear spin operator. We use the maximum curvature of each transition to estimate the lowest coherence time that can be calculated using the Eq. \ref{coherenceT}. Hence, $S_2$ is the largest of the absolute value of the eigenvalues of the $C$, and $S_1$ can be simply estimated using the magnitude of the $\nu$.

Energy levels given in Table.\ref{tab:dipole} can be considered as the ground states $\ket{g}$ and $\ket{s}$ as shown in Fig.\ref{fig: dark}.b.
Note that, because the dark state protocol is to some extent robust against the decoherence, the coherence time of the $\mathrm{^{167}Er}$ at zero field is long enough to achieve a high transfer fidelity and efficiency. 
Generally speaking, one can determine the required coherence time depending on the microwave cavity-spin coupling strength. To date, coupling of rare-earth spin ensembles to a microwave cavity has been demonstrated by several groups \cite{probst2013anisotropic, tkalvcec2014strong, staudt2012coupling, chen2016coupling}. For an efficient transfer, the cavity should be operated in the strong coupling regime and the system should remain coherent during the transfer time.

Transitions with zero gradient with respect to the magnetic field (so-called zero first-order Zeeman (ZEFOZ) points)  have a reduced sensitivity to the field fluctuations. Even isotopes of erbium have a pure first-order dependency on the magnetic field, and therefore, there is no ZEFOZ point for $^{168}$Er. However, an advantage of rare-earth ions that have an odd number of 4f electrons (i.e., Kramers ions) with non-zero nuclear spin (like $\mathrm{^{167}Er}$) is that the interaction between nuclear levels and electronic doublets can result in the ZEFOZ transitions even at zero magnetic field. 
For $\mathrm{^{167}Er}$, ZEFOZ points (where in Eq. (\ref{coherenceT}) $S_1=0$) at the zero field are associated with the transitions with sub-GHz frequencies. At this field, the longest coherence time we have estimated is 388$\mu s$ for the ZEFOZ transition frequency of 873 MHz. Although transitions of interest for the use in transducer protocols are those with frequencies of a few GHz, frequencies around $500$ MHz can still be used for interacting with fluxonium qubits \cite{nesterov2018microwave}. 


\section{CONCLUSION AND OUTLOOK}
\label{conclusion}
One of the important applications of quantum transducers is to connect quantum processors in a quantum network. In such a network, microwave-to-optical transducers can shift the wavelength of microwave photons to optical photons that are suitable for long distance quantum communications.
In this paper, using an optical and microwave cavity, we proposed the use of $\mathrm{^{167}Er}$:YSO as an intermediary for a microwave-to-optical quantum transducer.
 We presented a theoretical study of a proposed transducer design and calculated the achievable efficiency and fidelity of the system in the absence of external magnetic fields. Operating at nearly zero fields is important when interfacing with superconducting qubits. We have shown the robustness of the dark state protocol with respect to the dephasing rate. 
We then investigated ground state MW transitions and estimated transition frequencies, coherence times and transition strengths. The result can also be used in other transducer protocols where a detailed knowledge of the MW transitions is required. Note that, using the spin Hamiltonian parameters achieved from the crystal field model, one can also study the properties of the excited state energy levels.

Looking forward, our investigation on the MW transitions may provide further motivation for designing MW memories that interact with superconducting systems \cite{gouzien2021factoring}. 
In addition, the use of $\mathrm{^{167}Er}$:YSO has already been suggested for quantum repeaters \cite{asadi2020protocols}. Therefore, using $\mathrm{^{167}Er}$:YSO, one can think of an integrated system for wavelength transduction and long-distance entanglement distribution.

\section*{Acknowledgments}
The authors would like to thank Erhan Saglamyurek for useful discussions. This work was supported by the Natural Sciences and Engineering Research Council of Canada (NSERC) through its  Discovery  Grant and CREATE programs, the National Research Council (NRC) of Canada through its High-Throughput Secure Networks (HTSN) challenge program, and by Alberta Innovates Technology Futures (AITF).
\setcounter{equation}{0}
\setcounter{table}{0}
\setcounter{figure}{0}
\renewcommand{\theequation}{A\arabic{equation}}
\renewcommand{\thefigure}{A\arabic{figure}}
\renewcommand{\appendixname}{APPENDIX}
\appendix

\begin{figure}[!t]
    \includegraphics[scale=0.47]{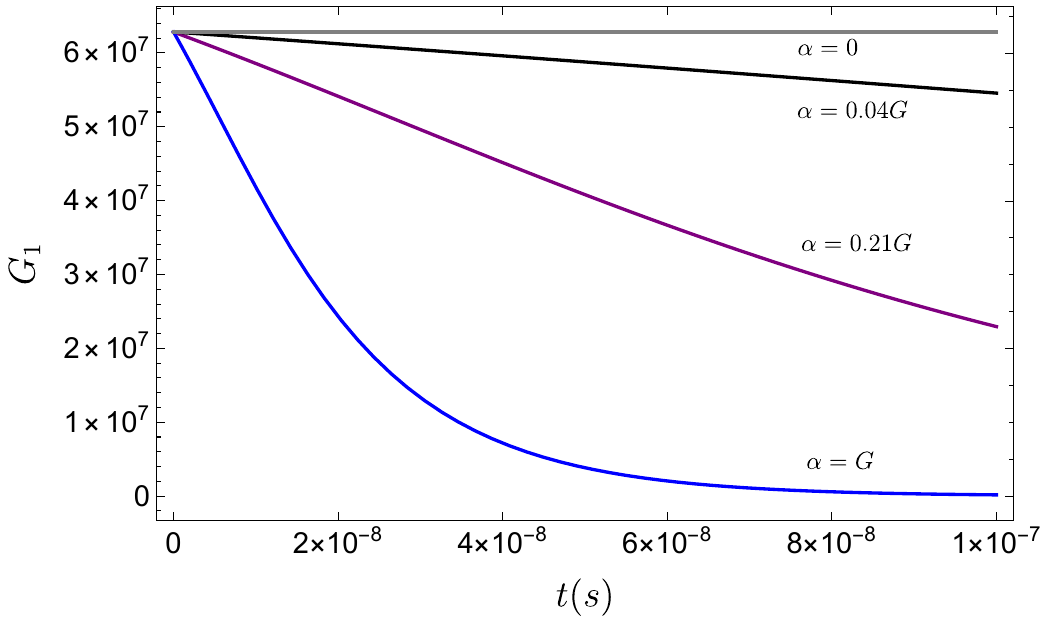}
	\caption{$G_1$ as a function of time for four different modulation strength parameters.} \label{fig:g1g2}
\end{figure}

\section{}
\subsection{The modulation for dark state protocol}
\label{Optimal-modulation}

As explained in the main text, for the modulation, we chose to set
$G_1(t)=G \sqrt{1-\text{tanh}(\alpha t)}$ and $G_2(t)=G$. To better understand the role of $\alpha$, in Fig.\ref{fig:g1g2}, we have plotted $G_1(t)$ for different values of $\alpha$. Here $\alpha=0$ is corresponding to $G_1(t)=G_2(t)=G$. In our case, we have set $\alpha= 0.212 G$.

\subsection{MW energy levels}
\label{energy-levels}
\begin{figure}[!t]
\includegraphics[scale=0.6]{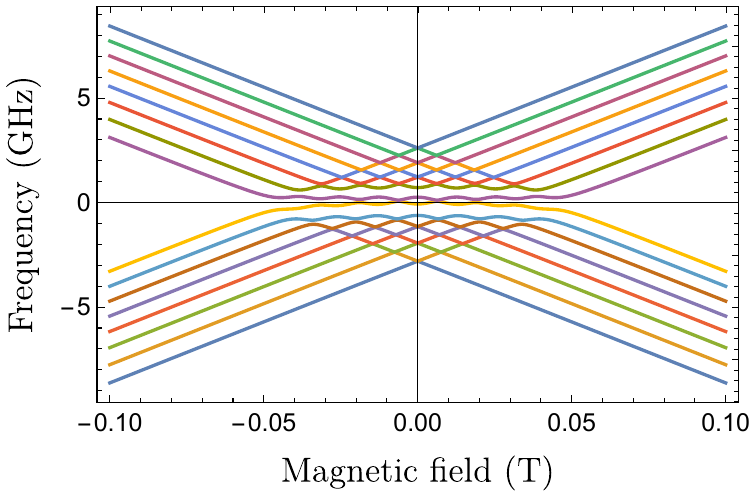}
	\caption{Ground state energy levels of the $\mathrm{^{167}Er}$:YSO as a function of external magnetic field along the $b$ axis.} \label{fig:baxis}
\end{figure}

In Fig. \ref{fig:baxis}, using the measured spin Hamiltonian parameters, we have plotted the $^{4}I_{15/2}$ ground state energy levels of
the $\mathrm{^{167}Er}$:YSO in the presence of a magnetic field along the $b$ axis.

			
			
			
	
	\label{tab:non-zero}



\FloatBarrier

	\bibliographystyle{apsrev4-1}
	\bibliography{ref}

\end{document}